\documentclass[prb,showpacs,preprint,a4paper]{revtex4}

\usepackage{amssymb}
\usepackage{graphicx}
\usepackage{bm}

\usepackage{epsfig}
\usepackage{amsfonts}
\usepackage{color}

\begin{document}

\title{Spectral properties and control of an exciton trapped in a multi-layered
quantum dot}
\author{Mariano Garagiola}
\email{mgaragiola@famaf.unc.edu.ar}
\author{Omar Osenda}
\email{osenda@famaf.unc.edu.ar}

\affiliation{Facultad de Matem\'atica, Astronom\'{\i}a y F\'{\i}sica,
Universidad Nacional de C\'ordoba, }
\affiliation{Instituto de F\'isica Enrique Gaviola, CONICET-UNC, Av. 
Medina Allende s/n, Ciudad Universitaria, X5000HUA C\'ordoba, Argentina }

\begin{abstract}
The spectral properties of one exciton trapped in a self-assembled multi-layered
quantum dot is obtained using a high precision variational numerical method.
The exciton Hamiltonian includes the effect of the polarization charges,
induced by the presence of the exciton in the quantum dot, at the material
interfaces. The method allows to implement rather easily the matching conditions
at the interfaces of the hetero-structure. The numerical method also provides
accurate approximate eigenfunctions that enable the study of the separability of
the exciton eigenfunction in electron and hole states. The separability, or the
entanglement content, of the total wave function allows a better understanding
of the spectral properties of the exciton and, in particular, shed some light
about when the  perturbation theory calculation of the spectrum is fairly
correct or not. Finally, using the approximate spectrum and eigenfunctions, the
controlled time evolution of the exciton wave function is analyzed when an
external driving field is applied to the system. It is found that it is
possible to obtain pico and sub-picoseconds controlled oscillations between two
particular states of the exciton with a rather low leakage of probability to
other exciton states, and with a simple pulse shape.
\end{abstract}
\date{\today}

\maketitle
\section{Introduction}\label{section:introduction}

The calculation of the spectral properties of an exciton trapped in a
semiconductor nano-structure is a long standing problem. Even in the case of
the most simple model used to study this system, {\em i.e.} the one
band effective mass approximation (EMA), there is a number of subtle details
that it is necessary to take into account. In particular, when the
nano-structure consists in a heterogeneous quantum dot, the mismatch between
the different material parameters imposes boundary conditions to the particle
wave functions and the electrostatic potential between the electron and hole
that form the exciton can not be taken as the simple Coulomb potential
\cite{Delerue2004,Ferreyra(1998)}. 

The presence of excitons can be easily spotted in the absorption spectrum 
of different semiconductors. The energy associated to each absorption peak is 
smaller to the energy difference between the electronic energy levels that 
lie on the semiconductor valence band and those that lie on the 
semiconductor conduction band. If the electrons in the semiconductor were
independent between them
and no many-body effects were present, the absorption energy should be equal to 
the energy difference between two electronic levels, one  located in the 
conduction band and the other located in the valence band. But, when an 
electron is promoted from the lower band to the upper one, by radiation
absorption for instance, the ``hole'' produced interacts 
with the electron, and the binding of the pair gives place to the exciton. The 
energy difference observed in the absorption spectrum, with respect to the 
values that correspond to independent electrons, is precisely given by the 
binding energy between the pair electron-hole
\cite{Cardona(2005),Bastard(1988)}. 

An exciton trapped in a quantum dot or in a given nano-structure has very
different physical properties compared to a bulk one. Roughly speaking, the
radius associated to  the exciton is, more or less, of the same magnitude order
than the characteristic length of many nano-structures. This can be used to tune
the physical trait of interest using the material parameters of
the semiconductors employed in the nano-structure, its geometry and its  size
 to tailor different properties. For instance, the mean
life time of the exciton can be adjusted depending on the application in mind.
Recently, some Quantum Information Processing (QIP) proposals use an exciton 
trapped
in a quantum dot as qubit,  in which case the basis states are given by the
presence ($\left| 1\right\rangle$), or absence ($\left| 0\right\rangle$), of
the
exciton \cite{Troiani(2000),Biolatti(2000)}. 
Obviously, for QIP applications the life time of the exciton
should be larger than the time needed to operate and control the qubit.

Modeling an exciton in a particular physical situation can be a tricky
business, since a plethora of effective Hamiltonians are available to use.
Starting from the multi-band many-electrons Hamiltonian, there is a number of
ways to derive effective Hamiltonians which usually decompose the one-electron
wave function in an ``envelop  part'' and a ``Bloch part''. The Bloch part
accounts for the underlying periodical lattice structure of the semiconductor
and the envelop part for the specific nano-structure under consideration. If
the electron is not confined, {\em i.e.} belongs to the bulk of the material,
the envelop part is reduced to the usual imaginary exponential present in the 
wave
functions allowed by the Bloch theorem. Otherwise, when a nano-structure is
present, the effective Hamiltonians for the envelop part resemble 
a multicomponent
Schr\"odinger-like equations. Probably, the most well know procedure to derive
effective-Hamiltonians is the {\bf k.p} method \cite{Voon(2009)}. The number of 
components of the
Hamiltonian derived using the  {\bf k.p} method depends on how much information 
about
the semiconductor band structure is incorporated. For instance, the well-known
eight-band model is derived using the fact that the valence band levels have
orbital angular momentum quantum number $L=1$, the conduction band levels $L=0$,
and the electronic spin has $s=1/2$ \cite{Haug(2004)}.

In this work we consider the simplest two-band effective mass approximation
model (EMA) for an exciton, trapped in a spherical Type -I device, with a core,
well, and barrier structure. The core and the barrier semiconductor compound is
exactly the same, while the well semiconductor is characterized by a conduction
band whose bottom energy is lower than the corresponding energy of the material
that form the core and the external barrier. The gap between the conduction
band and the valence band is narrower for the well semiconductor, than the gap
for the core one. Consequently, the confinement potential for the electron and
the hole are given by the profiles of the conduction and valence bands of the
semiconductors that form the device. This model has been studied extensively
because, despite its apparent simplicity, allows to obtain accurately the 
spectrum of different QD accurately in a wide variety  of situations. Anyway,
there are some features on it 
that ask
for some careful treatment. The first feature that must be handled with care is
owed to the model binding potential, since this is assumed as given by the
profile of the semiconductor bands, a step-like potential results for both, the
electron and the hole. Moreover, since the effective mass of both, the electron
and the hole, are assumed as discontinuous position dependent functions, the
Hermitian character of the kinetic energy is assured only with appropriate 
matching
conditions for the wave functions at the interfaces between different
materials. Finally, the electrostatic  potential between the electron
and hole can not be taken as the simple Coulomb potential. It has been shown
that polarization terms should be included because, again, of the presence of
interfaces between materials with different dielectric constants \cite{Ferreyra(1998)}.

It is worth to mention that the study of the properties of excitons confined in 
quantum dots 
\cite{Woggon(1995)} has been tackled using a number of methods, like 
perturbation theory \cite{Chang(1998),Schooss(1994)}, the {\bf k.p} 
method \cite{Pokatilov(2001),Efros(1998)}, and with different types of 
confinement potential as the infinite potential well \cite{Ferreyra(1998)} or 
the parabolic one \cite{Garm(1996)}. The selection of potentials and method are 
often  dictated by the application in mind, that could range from excitonic 
lasers \cite{Bimberg(1997),Bimberg(2005)}, through quantum information 
processing \cite{Kamada(2004),Chen(2001),Biolatti(2002)}, up to 
one-electron transistors \cite{Ishibashi(2003)} in micro electronics devices. 
Between the physical phenomena that have received more attention, it can be 
mentioned the binding energy \cite{Lelong(1996)}, the decoherence 
\cite{Calarco(2003),Bonadeo(1998)A,Bonadeo(1998)B} and the Stark 
effect \cite{Billaud(2009)} among many others.

To study the two-band EMA model described above we calculated its spectrum and
eigen-states using a variational approach. The approach allows us to take into
account the matching conditions at the interfaces of the device and the (quite)
complicated interaction potential between the two parts of the exciton that
includes the effects of polarization. As we shall see, our approach allows us to
obtain highly accurate approximate results for the spectrum and approximate
wave functions that provide information about the entanglement content of the
two-particle quantum state. This information gives useful information about the
parameter range where the two-particle wave function can be more or
less accurately taken as the product of one-particle wave functions.
The manuscript is organized as follows: in Section~\ref{section:one-body-model} 
the variational method to obtain the spectrum of one or two-particle 
Hamiltonians is discussed and presented in some detail, In 
Section~\ref{section:two-particle} the electron-hole pair EMA Hamiltonian is 
analyzed paying some attention to the electrostatic problem originated by the 
geometry and composition of the quantum dot and the binding energy of the 
electron-hole pair is calculated, in Section~\ref{section:separability} the 
separability problem of the exciton wave function is presented and analyzed, in 
particular it is shown that the characterization through a measure of 
separability gives a better understanding of the exciton spectral properties. 
The time evolution of the exciton wave function when an external driving field 
is applied to the QD is studied in Section~\ref{section:control}. The study is 
intended to look for regimes where the external driving allows to switch between 
only two pre-selected states of the exciton, with a negligible or controllable 
leakage of probability to other exciton states. Finally, our results are 
discussed and summarized in Section~\ref{section:conclusions}

\section{One body models and methods}\label{section:one-body-model}

The EMA approximation, when applied to the description of an independent
electron (e), or hole (h), is characterized by a number of parameters such as
the effective mass of the particle or the energy band gap of the materials from
which is made the quantum dot. Otherwise, the Hamiltonian associated to it seems
like an ordinary Schr\"odinger-like Hamiltonian
\begin{equation}\label{eq:one-body-ham}
\mathcal{H}_{e(h)}=-\frac{\hbar^2}{2}\nabla_{e(h)}\left(\frac{1}{m_{e(h)}^*(r_{
e(h)})}\nabla_{e(h)}\right)+
V_{e(h)}(r_{e(h)})+V_{s}(r_{e(h)})
\end{equation}
where 
$m_{e(h)}^*(r_{e(h)})$ is the electron (hole) effective mass. For an one-band
model, $m_{e}^*$ corresponds to the the effective mass of the electron in the
conduction band, while $m_{h}^*$ is taken as the {\em light} effective mass of
the hole in the valence band. $V_{e(h)}(r_{e(h)})$ is the binding potential for
the electron (hole).

\begin{figure}[floatfix]
\begin{center}
\includegraphics[scale=0.3]{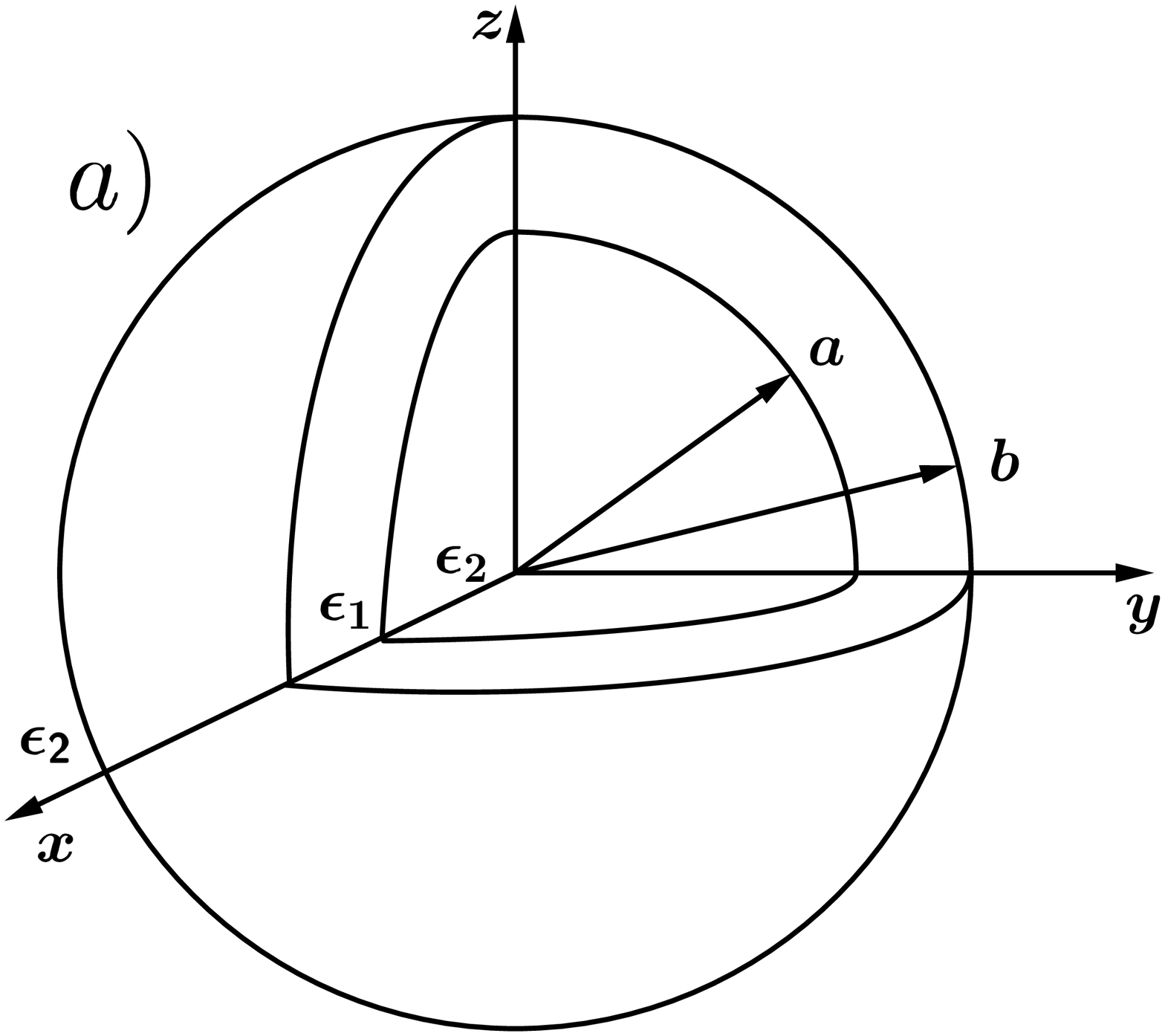}
\includegraphics[scale=0.9]{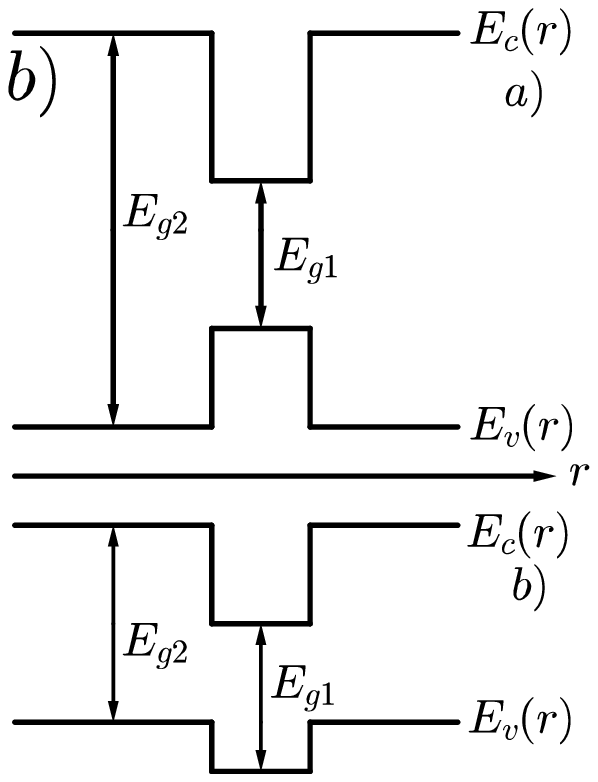}
\end{center}
\caption{\label{fig:tipoQD} a) The cartoon shows the structure of the spherical
self-assembled quantum dot. The different layers correspond to, from the center
and going in the radial direction,  the core whose radius $a$ and dielectric
constant are shown, the potential well, with inner radius $a$ and exterior
radius $b$, and the clad. b) The profiles of the conduction band, $E_c(r)$, and
of the valence band, $E_v(r)$, for both a Type-I device (top) and a a Type-II
devive (bottom). $E_g1$ and $E_g2$ are the gap energies of the potential well
material and the core/clad material, respectively.}
\end{figure}

Figure \ref{fig:tipoQD}a) shows a cartoon of the spherical self assembled
quantum dot under consideration. The inner core, of radius $a$, and the outer
shell (also called clad) are made of the same compound, let us call it two (2),
while the middle layer, of radius $b$,  is made of a different compound, let us
call it one (1). These kind of hetero structures are termed of Type I or II, if
the top energy of the valence band of the middle layer material, $E^1_{top}$,
is larger than the top energy of the valence band of material two, $E^2_{top}$,
{\em i.e.} the hetero structure is of Type I if $E^1_{top} > E^2_{top}$,
otherwise is of Type II, as can be seen clearly in Figure \ref{fig:tipoQD}
b).
The cartoon in Figure~\ref{fig:tipoQD}b) shows the conduction and valence bands
profile as a function of the
radial distance from the center of the hetero-structure.

Since the potential in Equation \ref{eq:one-body-ham}, $V_{e(h)}(r_{e(h)})$, is
taken as exactly the conduction band profile  (valence band profile) for the
electron (hole), it is clear that a Type-I device corresponds to the case where
the middle layer acts as a potential well for both particles, so
\begin{equation}
\label{eq:potencial}
V_{e(h)}(r)=\left\{\begin{array}{lcl}
V_0^{e(h)}>0& &0<r<a\\
0& &a<r<b\\
V_0^{e(h)}>0& &b<r
\end{array}\right.,
\end{equation}
and $V_0^{e(h)}=const.$

The potential
$V_s$ is owed to the polarization charges produced at the interfaces
core/middle layer and middle layer/clad because of the mismatch between their
dielectric constants. It can be shown that the auto-polarization potential is
given by
\begin{equation}
\label{eq:polarizacion}
V_s(r)=\frac{q^2}{2\,
\varepsilon_1}\sum_{l}\frac{1}{(1-pq)}\left(qr^{2l}+\frac{p}{r^{2l+1}}+\frac{pq}
{r}
\right)
\end{equation}
where
\begin{equation}
p=(\varepsilon_1-\varepsilon_2)la^{(2l+1)}/[\varepsilon_2 l+\varepsilon_1 
(l+1)],
\end{equation}
and 
\begin{equation}
q=(\varepsilon_1-\varepsilon_2)(l+1)b^{-(2l+1)}/[\varepsilon_1 l+\varepsilon_2 
(l+1)] .
\end{equation}

The kinetic energy term in Equation \ref{eq:one-body-ham} is written in a
symmetric fashion in order to guarantee the Hermitian character of the
operator, if it is considered together with the following matching conditions

\begin{eqnarray}
\label{eq:matching}
\psi(r=a^-)&=&\psi(r=a^+) \\ \nonumber
\psi(r=b^-)&=&\psi(r=b^+) \\ \nonumber
\left( \frac{1}{m^*} \frac{d\psi}{dr}\right) |_{r=a^-} &= &\left( \frac{1}{m^*}
\frac{d\psi}{dr}\right) |_{r=a^+} \\ \nonumber
\left( \frac{1}{m^*} \frac{d\psi}{dr}\right) |_{r=b^-} &= &\left( \frac{1}{m^*}
\frac{d\psi}{dr}\right) |_{r=b^+} .
\end{eqnarray}

The eigenvalue problem 
\begin{equation}
\label{eq:eigenvalue-problem}
\mathcal{H} \psi = E \psi ,
\end{equation}
with $\mathcal{H}$ given by Equation~\ref{eq:one-body-ham} can be studied using
different (an numerous) methods that result in an approximate spectrum and, in
some cases, approximate eigenfunctions. Anyway, the matching conditions can
not be implemented in a direct way depending on which method is used to tackle
the problem. A method that allows to weight adequately the matching conditions
is one that incorporates easily the step-like nature of the binding potential,
Equation~\ref{eq:potencial}, and the effective mass. The
approximate eigenfunctions and eigenvalues analyzed in this work were obtained
using B-splines basis sets, which have been used to obtain high accuracy
results for atomic, molecular and quantum dot systems. The method has been well
described elsewhere, so the only details that we discus here are the relevant
ones to understand some of results presented later on.

To use the B-splines basis, the  normalized one-electron
orbitals are given by
\begin{equation}\label{phi-bs}
\phi_{n}({r}) = C_n \, \frac{B^{(k)}_{n+1}(r)}{r}  \,;\;\;n=1,\ldots
\end{equation}

\noindent where $B^{(k)}_{n+1}(r)$ is a B-splines polynomial of order $k$.
The numerical results  are obtained by defining a cutoff radius $R$, and
then the interval $[0,R]$ is divided into $I$ equal subintervals.
 B-spline polynomials \cite{deboor} (for a review
of applications of B-splines polynomials in atomic and molecular physics,
see Reference \cite{bachau01}, for its application to QD problems see Reference
\cite{Ferron2013}) are piecewise polynomials defined by a
sequence of  knots $t_1=0\leq t_2\leq\cdots \leq t_{2 k+I-1}=R$
and the recurrence relations

\begin{equation}\label{bs1}
B_{i}^{(1)}(r)\,=\,\left\{ \begin{array}{ll} 1 & \mbox{if}\,t_i\leq r <
t_{i+1}   \\
0 &\mbox{otherwise,}  \end{array}  \right. \,.
\end{equation}

\begin{equation}\label{bsrr}
B_{i}^{(k)}(r)\,=\,\frac{r-t_i}{t_{i+k-1}-t_i}\,B_{i}^{(k-1)}(r)\,+\,
\frac{t_{i+k}-r}{t_{i+k}-t_{i+1}}\,B_{i}^{(k-1)}(r)\; (k>1)\,.
\end{equation}

The standard  choice for the knots  in
atomic physics  \cite{bachau01} is $t_1=\cdots=t_k=0$ and
$t_{k+I}=\cdots=t_{2k+I-1}=R$. Because of the matching conditions at the
interfaces between the core and the potential well and between the potential
well and the clad, it is more appropriate to choose the knots as follows: at
the extremes of the interval $\left[0,R\right]$ where the wave function is
calculated, there are $k$ knots that are repeated, while at the interfaces
$r=a$ and $r=b$ there are $k-3$ knots that are repeated, and in the open
intervals $(0,a)$, $(a,b)$ and $(b,R)$ the knots are distributed uniformly
\cite{HaoXue(2002)}

The constant $C_n$ in Equation~\ref{phi-bs} is a normalization constant obtained
from the condition $\left\langle \phi_n |\phi_n \right\rangle =1 $

\begin{equation}\label{nor-c}
C_n = \frac{1}{\left[ \int_0^{R} \, \left(B^{(k)}_{n+1}(r) \right)^2 \,dr
\right]^{1/2}} \,.
\end{equation}

Because $B_1(0)\ne0$ and $B_{I+3k-9}(R)\ne0$, we have $N=I+3k-11$ orbitals
corresponding to $B_2,\ldots,B_{I+3k-10}$. In all the calculations we used
the value $k=5$, and, we do not write the index $k$ in the eigenvalues and
coefficients.

\begin{figure}[floatfix]
\begin{center}
\includegraphics[scale=0.6]{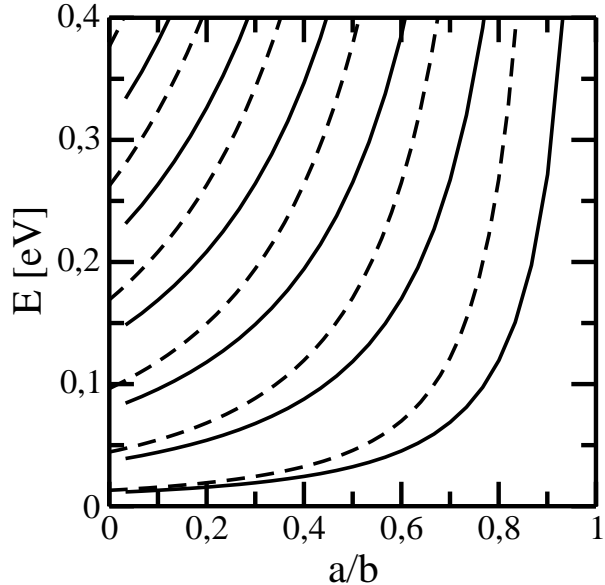}
\end{center}
\caption{\label{fig:fin-vs-inf} The first one-electron eigenvalues for a 
infinite potential well (dashed lines) and for a finite potential well (solid 
lines) as functions of the dimensionless ratio $a/b$. In both cases, $b=31.71$ 
nm, that is the Bohr radius for an exciton immersed in a HgS matrix. The 
eigenvalues correspond to zero angular momentum ($\ell=0$) states.}
\end{figure}

To gain some insight about the performance of the 
B-spline method we first studied the eigenvalue problem of  one electron
confined in a multi-layered quantum dot as the one depicted in
Figure~\ref{fig:tipoQD}a). The quantum dots whose core, well and clad are made
of CdS/HgS/CdS have been extensively studied \cite{Schooss(1994)}, so it is a
good starting point to
our study.
For these materials, the effective masses
for the electron in their respective conduction bands are $m_{e,CdS}^*=0.2$,
$m_{e,HgS}^*=0.036$, while for the hole in the valence bands the effective
masses are given by $m_{h,CdS}^*=0.7$, $m_{h,HgS}^*=0.040$. The dielectric
constants are $\varepsilon_{CdS}=\varepsilon_2=5.5$,
$\varepsilon_{HgS}=\varepsilon_1=11.36$. Besides,
\begin{eqnarray*}
\label{eq:band-energies}
E^{CdS}_{bottom} - E^{HgS}_{bottom} & = & 1.35\mbox{eV}, \\ 
E^{HgS}_{top} - E^{CdS}_{top} & = & 0.9\mbox{eV} .
\end{eqnarray*}
The other parameters that define the device are the radii $a$ and $b$. To
augment the effect of the confinement owed to the binding potential we choose
$b$ equal to the Bohr radius of a bulk exciton in HgS \cite{Chang(1998)},
then $a$ can take any value between zero and $b$. This model was studied in
\cite{Ferreyra(1998)} considering the limit of ``strong confinement'', {\em
i.e.} the electron is bound in an infinite potential well. We want to remark
that the electro-hole pair behaves very much as a hydrogen-like atom when it is
immersed in a semiconductor bulk besides, in many cases, the hole
effective mass is much larger than electron one. So, it makes sense to consider
the Bohr radius as one of the length scales that characterize the problem.

Figure~\ref{fig:fin-vs-inf} shows the behavior of the lowest lying approximate
one-electron eigenvalues obtained using the B-spline method for a quantum dot
with the material parameters listed
above, and for one electron bound in an infinite potential well, as functions
of the ratio between both radii, $a/b$. The eigenvalues corresponds to
eigenfunctions with orbital angular momentum with quantum number $\ell=0$. As
can be observed from the Figure, the infinite potential well eigenvalues, that
can be obtained exactly, are a pretty good approximation to the quantum dot
eigenvalues for small values of $a/b$, but for larger values of $a/b$, or for
the excited states, the relative error grows considerably. Since in many works
in the literature, the binding energy of the exciton is obtained using the exact
solutions of the infinite potential well, the results shown in
Figure~\ref{fig:fin-vs-inf} indicate that  it is necessary to proceed with great
caution if this approximation were to be used. The eigenvalues obtained using
the B-spline method have a relative error of less than $10^{-3}$.

\begin{figure}[floatfix]
\begin{center}
\includegraphics[scale=0.6]{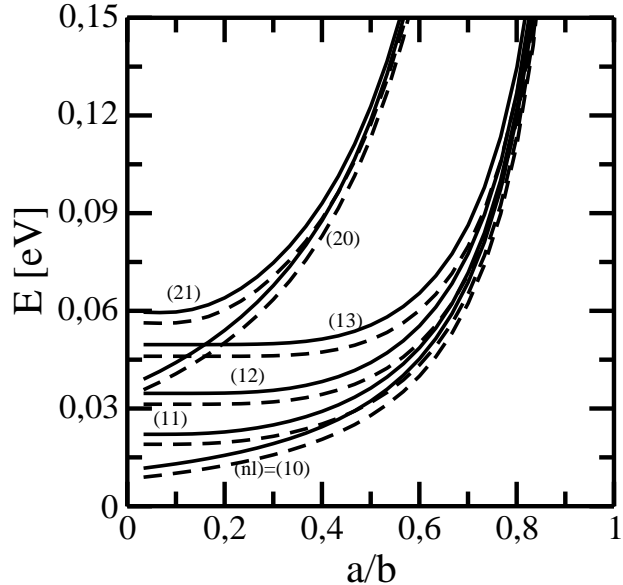}
\end{center}
\caption{\label{fig:auto-polarization} The lowest lying one-electro eigenvalues 
for different radial ($n_r$) and angular momentum ($\ell$) quantum numbers. The 
eigenvalues were calculated for a finite potential well and the Hamiltonian 
includes the auto-polarization term in Equation~\ref{eq:polarizacion} (solid 
lines). The dashed lines are the corresponding eigenvalues obtained taking out 
the autopolarization term (or equivalently choosing 
$\varepsilon_1=\varepsilon_2$). The labels for each curve are the  quantum 
numbers that identify the corresponding eigenstate.
}
\end{figure}

There has been some discussion about the need to include, or not, the
self polarization term, Equation~\ref{eq:polarizacion}, in the one
electron Hamiltonian, Equation~\ref{eq:one-body-ham} \cite{Delerue2004}.
Figure~\ref{fig:auto-polarization} shows the behaviour of the lowest
lying energy eigenvalues, for several orbital angular momentum quantum numbers,
and with, or without, the auto-polarization term. Is is clear that for
$a/b<0.4$, and for all the angular momentum quantum numbers shown, that the
auto-polarization term changes the eigenvalues in a, approximately, 10$\%$ ,
showing
that for large quantum dot the auto-polarization contribution can not be
ignored. When $a$ grows up to
$b$ the effect becomes less and less important. Other way to characterize the
phenomenology is the following, if the eigenvalues become independent of $\ell$
for a given radial quantum number $n$, {\em i.e.} the energy eigenvalues
depends almost only on $n$, then the well potential becomes ``small enough'' and
the auto-polarization term becomes negligible. In this limit, the polarization
term and the angular momentum part of the kinetic energy can be treated as
perturbations.

It is interesting to point that despite that the hole Hamiltonian is determined
by different parameters than the electron one, the behavior of its spectrum is
very similar, for that reason we do not include a detailed analysis of it. On
the other hand, once the electron and hole approximate eigenfunctions are
obtained, a plot of them reveals that both particles are well localized inside
the potential well. Besides,  the repetition of knots at the interfaces enables
that the approximate solutions meet the matching
conditions, Equation~\ref{eq:matching}, to a high degree of accuracy.

\section{Two-particle model}\label{section:two-particle}

The Hamiltonian for an exciton formed by an electron and a hole can be written
as 
\begin{equation}
\label{eq:hamiltoniano-exciton}
\mathcal{H}_{ex}=\mathcal{H}_{e}+\mathcal{H}_{h}+V_c({\mathbf
r}_e,{\mathbf r}_h) ,
\end{equation}
where $\mathcal{H}_{e}$ and $\mathcal{H}_{h}$ are the one-body Hamiltonians of
Equation~\ref{eq:one-body-ham}, and $V_c({\mathbf r}_e,{\mathbf r}_h)$ is the
electrostatic potential between the electron and hole. Usually, one is tempted
to consider $V_c$ as the usual Coulomb potential between a positive charge and a
negative one but, if the exciton is confined to an hetero-structure made of
different materials, this approach oversimplifies the situation.

Having in mind the argument of the paragraph above, we consider a better a
approximation for the actual electrostatic potential that was suggested by
Ferreyra and Proetto \cite{Ferreyra(1998)}. Since the hole is usually ``heavier''
than the electron, and that the scenario of most interest occurs when both
particle are well localized, it is simpler to consider $V_c$ as the solution to
the Poisson equation considering that the hole and electron coordinates are
restricted to  the potential well, {\em i.e.}
\begin{equation}
V_c({\mathbf r}_e,{\mathbf r}_h) = q_e q_h \, G({\mathbf r}_e,{\mathbf r}_h) ,
\end{equation}
where
\begin{equation}\label{eq:green}
\nabla_r^2 G({\mathbf r},{\mathbf r}') = \left\{
\begin{array}{lcl}
-\frac{4\pi}{\varepsilon_1} \delta({\mathbf r}-{\mathbf r}') & \quad & 
\mbox{if}\;
a<r<b \\
0 & & \mbox{otherwise}
\end{array}
\right. ,
\end{equation}

\begin{figure}[floatfix]
\begin{center}
\includegraphics*[scale=0.6]{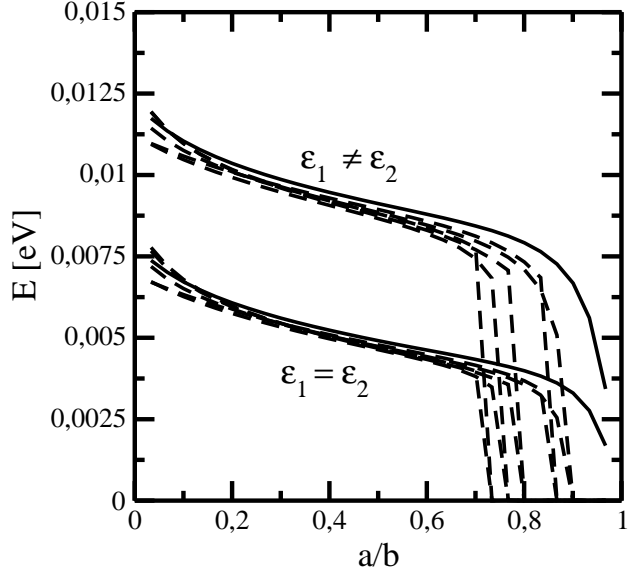}
\end{center}
\caption{\label{fig:binding-energy} The binding energy expectation value as a 
function of the ratio $a/b$. The curves are grouped in two bundles, the higher 
one was calculated including all the polarization effects originated by the 
dielectric constants mismatch, $\varepsilon_1\neq\varepsilon_2$ , while the 
lower one ignores the mismatch and takes $\varepsilon_1 = \varepsilon_2$. In 
each bundle, the curve at the top (solid line) corresponds to the exciton 
ground state. As the binding energy is a decreasing function of the exciton 
eigen-energy, in each bundle the lower and lower lying curves correspond to the 
first excited state, to the second one, ans so successively. The sudden drop of 
the curves observed for large enough values of the ratio $a/b$ show where each 
level reach the ``ionization threshold''.}
\end{figure}

So, solving Equation~\ref{eq:green}, it can be shown that the electrostatic
potential that gives the interaction between the electron
and the hole can be written as

\begin{eqnarray}
\label{eq:electrostatic-potential}
V_c(\mathbf{r}_e,\mathbf{r}_h)&=&q_eq_h\sum_{l,m}Y_{lm}^*(\theta_h,\varphi_h)Y_{
lm }
(\theta_e,\varphi_e) \\ \nonumber
                       & &\quad \times \frac{4\pi}{\varepsilon_1(2l+1)(1-pq)} \\
                        &&\quad \times [r_<^l+pr_<^{l+1}][r_>^{-(l+1)}+qr_>^l] ,
\end{eqnarray}
where
\begin{equation}\label{eq:p-term}
 p=(\varepsilon_1-\varepsilon_2)la^{(2l+1)}/[\varepsilon_2 l+\varepsilon_1
(l+1)] ,
\end{equation} 
\begin{equation}\label{eq:q-term}
q=(\varepsilon_1-\varepsilon_2)(l+1)b^{-(2l+1)}/[\varepsilon_1 l+\varepsilon_2 
(l+1)] ,
\end{equation}
and
\begin{equation}  
r_{>(<)}=max(min) \{r_e, r_h\}.
\end{equation}

The exciton binding energy can be obtained as the expectation value of the
electrostatic potential
\begin{equation}
\label{eq:binding-energy}
E_{binding}^{\alpha}= -\langle
\psi_{\alpha}|V_c(\mathbf{r}_e,\mathbf{r}_h)|\psi_{\alpha}\rangle,
\end{equation}
where $|\psi_{\alpha}\rangle$ is an eigenstate of the exciton Hamiltonian,
Equation~\ref{eq:hamiltoniano-exciton}.

Figure~\ref{fig:binding-energy} shows the behavior of the binding energy as a
function of the radii ratio $a/b$. For clarity, we restrict the curves shown to
those data obtained with eigenfunctions with orbital angular momentum quantum
numbers $l_e=l_h=m_e=m_h=0$. The Figure shows two well defined separated sets
of curves. The lower set of curves correspond to the binding energy calculated
without polarization terms (or equivalently putting 
$\varepsilon_1=\varepsilon_2$ in
Equations~\ref{eq:electrostatic-potential},\ref{eq:p-term} and
\ref{eq:q-term}). The upper set of curves corresponds to the binding energy
calculated considering all the polarization effects ($\varepsilon_1\neq 
\varepsilon_2$).
The polarization terms, owed to the polarization charges at the interfaces
between the different materials, does not change the qualitative behavior of
the binding energy but, at least for the parameters of
Figure~\ref{fig:binding-energy}, not including them leads to an
underestimation of the binding energy of approximately 100\% . The
inclusion of the polarization terms in the electrostatic potential increases
the binding energy since for $\varepsilon_1>\varepsilon_2$ the correction terms 
in
Equation~\ref{eq:electrostatic-potential}, with respect to the Coulomb
potential, have all the same sign.

\section{separability of the exciton
eigenfunctions}\label{section:separability}

The availability of accurate numerical
approximations for the actual exciton eigenfunctions gives the possibility to
analyze their separability, in particular in this Section we analyze the
behavior of the von Neumann entropy associated to the excitonic quantum state.
We will show that some features of the binding energy already described in the
precedent Section, can be understood best studying simultaneously both
quantities. In particular, the analysis of the separability of the quantum
states can shed some light about when the interaction between the exciton
components can be treated using perturbation theory. Besides, there are some
quantities that determine the strength of the interaction of the exciton with
external fields, as the dipole moment, or expectation values that can not be
accurately obtained if the correlation between the electron and hole is not
taken into account.

A well known separability measure of the quantum state is the von Neumann
entropy, $S$, which can be calculated as
\begin{equation}\label{eq:von-Neumann-def}
S=-\sum_k \lambda_k \; \ln(\lambda_k) ,
\end{equation}
where $\lambda_k$ are the eigenvalues of the electron reduced density matrix
$\rho^{red}(\mathbf{r}_e,\mathbf{r}^{\prime}_e)$ which, if the quantum state of
the exciton is given by a vector state $\psi(\mathbf{r}_e,\mathbf{r}_h)$, then
\begin{equation}\label{eq:reduced-density}
\rho^{red}(\mathbf{r}_e,\mathbf{r}_e^{\prime} ) = \int
\psi^{\star}(\mathbf{r}_e,\mathbf{r}_h) \psi(\mathbf{r}_e',\mathbf{r}_h) \;
d\mathbf{r}_h .
\end{equation}

\begin{figure}[floatfix]
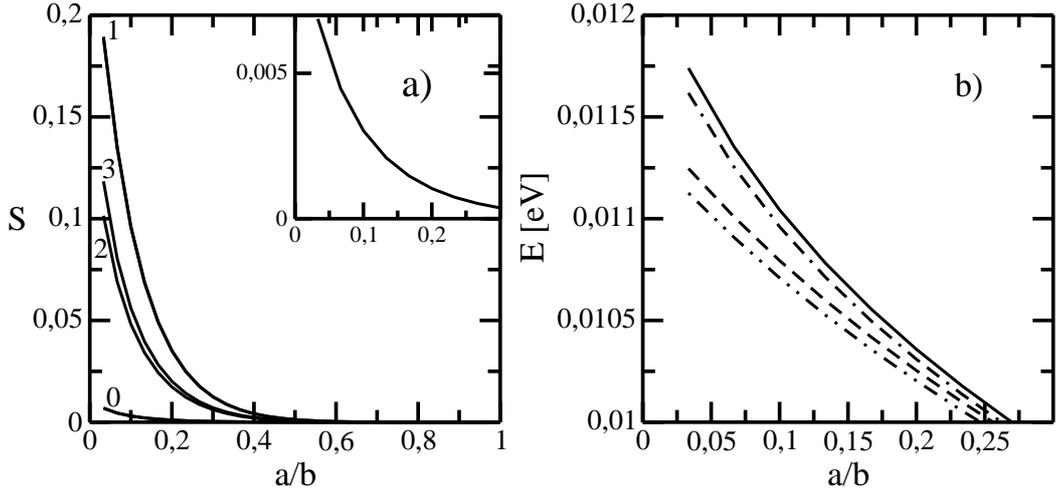

\begin{center}
\includegraphics[scale=0.5]{fig7.eps}
\includegraphics[scale=0.5]{fig8.eps}
\end{center}
\caption{\label{fig:von-Neumann-entropy} a) The von Neummann entropy for the 
first few eigenstates, the lowest curve 
corresponds to the ground state eigenstate (label 0), the following to the 
second excited eigenstate (label 2) and so on. The inset 
shows a detailed view of the fundamental state von Neumann entropy. b) The 
binding energy for the ground state calculated following different methods. 
From top to bottom the curves were obtained with the full Hamiltonian and the 
B-spline method (solid line), without the auto-polarization terms ({\em i.e.}
neglecting $V_s(r)$) and the 
B-spline method (dot-dashed line), again with the full Hamiltonian but using 
perturbation theory (dashed line)  and finally without the auto-polarization
terms 
and using perturbation theory (double-dot dashed line), respectively. For 
details see the text }
\end{figure}

Figure~\ref{fig:von-Neumann-entropy}a shows the behavior of the von Neumann
entropy as a function of the ratio $a/b$, for the approximate eigenfunctions of
the first
low lying exciton eigenvalues calculated using the B-spline method. As can be
appreciated from the figure, the von
Neumann entropy is a monotone decreasing function of the ratio $a/b$, {\em
i.e.} the electron and hole became more and more independent (separable its
wave function) when the radius of the core is increased. This is expected, but
even for the ground state the effect of the non-separability of the excitonic
wave function has a rather large influence in the binding energy, as can been
appreciated in panel b). The von Neumann entropy, on the other hand is 
larger for the exciton excited states so, at least in principle, any effect
related to the non-separability of the exciton wave function should be
stronger for the excited states.

Figure~\ref{fig:von-Neumann-entropy}b shows the
behaviour of the groud state binding energy, again as a function of the ratio
$a/b$, obtained using different methods. The top curve corresponds to the
binding energy calculated with the B-spline approximation while, in decreasing
order, the figure also shows the curves that correspond to the binding energy
calculated with the B-spline method but without including the auto-polarization
terms, besides the binding energy calculated using perturbation theory with
and without
the auto-polarization terms. The binding energy curve obtained using
perturbation theory shows the first order approximation energy calculated using
the finite potential well electron and hole eigenfunctions as the unperturbed
levels. 

At least for this set of parameters and materials, taking into account
the auto-polarization terms has less influence than using a method (the
B-splines) that takes into account the non-separability of the two-particle
wave function. For the ground state the worst case scenario, perturbation
theory without auto-polarization, differs from the best one, B-splines plus
auto-polarization terms, by about five percent. Of course, since the
self-assembled quantum dots offer a huge amount of different combinations of
materials, geometries and sizes the quantitative results may change more or
less broadly.

\begin{figure}
\begin{center}
\includegraphics[scale=0.6]{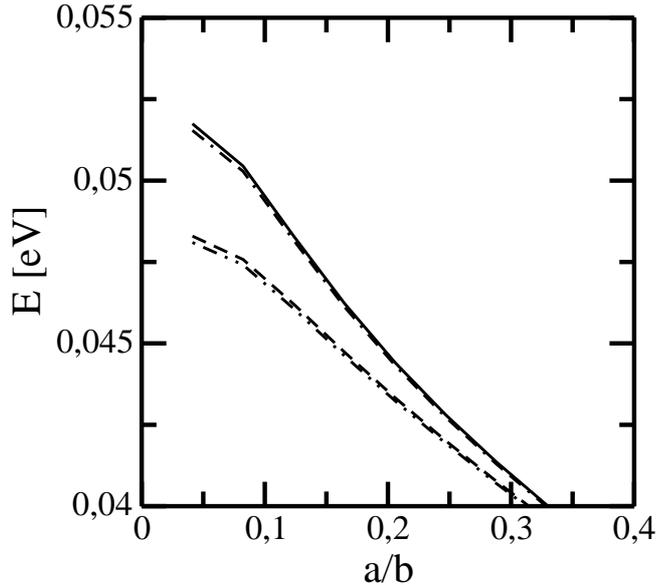}
\end{center}
\caption{\label{fig:otro-dot} The ground state exciton binding energy {\em vs} 
$a/b$ for a device with core,well and clad made of ZnS/CdSe/ZnS. The curves are 
obtained following the same prescription than those shown in 
Figure~\ref{fig:von-Neumann-entropy}. All the materials parameters can be found
in Reference~\cite{Chang(1998)}. It is easy to appreciate that for this device
the influence of the 
auto-polarization terms is smaller that in the first device analyzed, but the 
difference between the values obtained using perturbation theory and the 
B-spline method is larger}
\end{figure}

Figure~\ref{fig:otro-dot} shows the binding energy obtained with
the  same methods  that were used to obtain the data in
Figure~\ref{fig:von-Neumann-entropy}b), as describe above, but for an exciton
in a different quantum dot. In this case, we considered a quantum dot formed by
core/well/barrier structure made of ZnS/CdSe/ZnS. All the necessary parameters
to determine  the Hamiltonian, effective masses, dielectric
constants, etc,  can be found in Reference~\cite{Schooss(1994)}.

For the  case shown in Figure~\ref{fig:otro-dot}, the influence of the
auto-polarization terms is even smaller that in the first case analyzed,
Figure~\ref{fig:von-Neumann-entropy}, and the difference between the best and
worst scenarios defined above is around seven, eight percent.

\section{control}\label{section:control}

There are two important applications of excitons in which the speed with which
you can go from having to not having a exciton is crucial: the controlled and
rapid photon production and the switching between the basis states of the qubit
logically associated to the exciton. The physical process is exactly the same,
but the motivation and requirements depend on which application is being
considered. In this section we are interested in estimating how much we can
force an excitonic qubit so that it oscillates periodically between its basis
states with the highest possible frequency and  a small  probability loss 
to other exciton eigenstates. The need to estimate how fast the qubit can be
switched between its basis states comes from the limits imposed by the
decoherence sources that are unavoidable and restrict seriously the total
operation time in which the qubit would keep its quantum coherence. Moreover, we
intend to achieve the switching between the basis states with the simplest
control pulse, that is, a sinusoidal one. So, many times it is found reasonable
to ask that, if $T_S$ is the switching time and $\tau$ the  time scale
associated to the decoherence processes present in the physical system, then
$T_S\sim 10^{-4} \tau$.

The main source of decoherence in charge qubits is owed to the interaction of
the charge carrier, the electron(s) trapped in the quantum dot, and the thermal
phonon bath present in the semiconductor matrix. This is the reason why, in many
cases, spin qubits  are preferred although its control is more
complicated and have more longer operation time. Since in the exciton case the
strength of the coupling with the phonon bath depends on the difference between
the
electronic and hole wave functions, it is to be expected that the decoherence
rate for qubits based on one exciton would be smaller than for a charge qubit
based on one (or more) electrons trapped in a quantum dot. Ideally, when the
potential wells for the electron and the hole have exactly the same depth, 
for equal effective masses, and in the limit of zero interaction, the coupling
with the phonon bath almost disappears. In this sense, the separability of the
electron-hole wave function provides a good measure to select the parameter
region where the coupling of the exciton with the phonon bath is smaller. 
Consequently, since even for very low temperatures the decoherence produced
by the phonon bath imposes a total operation time on the order of a
few tens of nanoseconds for qubits based on multi-layered self-assembled quantum
dots, then to be considered a putative useful qubit the switching time $T_S$
should be on the order of the picoseconds or sub-picoseconds. 

The leaking of probability to other exciton eigenstates  when an external
driving is applied can be analyzed using the following unitary one-exciton
Hamiltonian, which describes the interaction of the
electron-hole dipolar moment, $\vec{d}$, with an external periodic field
$\vec{E}(t)$
applied to the quantum dot \cite{Haug(2004)}
\begin{equation}
\mathcal{H}_{int}(t)=-\vec{d}\ldotp\vec{E}(t) ,
\end{equation}
which in second quantization formalism can be written as
\cite{Haug(2004),Biolatti(2002)}
\begin{equation}
\label{interaccion}
\mathcal{H}_{int}(t)=-E(t)\sum_{nm} \left[ \mu_{nm}^* a_{n}^{\dagger}
b_{m}^{\dagger} + h.c\right] ,
\end{equation}
where $a^{\dagger}_n$  is the creation operator of an electron in the conduction
band, $b^{\dagger}_m$ is the creation operator of a hole in the valence band,
$n$ and $m$ stand for the corresponding one-particle levels,
and $\mu^{\star}_{nm}$ is the matrix element of the dipolar moment operator,
given by
\begin{equation}
\mu_{nm}=\mu_{bulk}\int \phi_{n}^e(\vec{r})\phi_{m}^h(\vec{r})d^3r, 
\end{equation}
where $\mu_{bulk} $ is the dipolar moment corresponding to an electronic
transition from the valence band to the conduction
band\cite{Haug(2004),Biolatti(2002)} . 

All the one- and two-particles quantities needed to determine the parameters in
Equation~\ref{interaccion} and the time evolution of the exciton state can be
obtained using the B-spline method described in the preceding Sections.
Moreover,
since the B-spline method provides a very good approximation to all the
exciton bound states the time evolution of the approximate exciton quantum state
can be
written as a sum over bound states as follows
\begin{equation}
\Psi(t) = \sum_i U_i(t) \psi_i ,
\end{equation}
where the sum runs over the approximate bound states provided by the B-spline
method,  $\psi_i$, 
and the time-dependent coefficients $U_(t)$ can be calculated integrating a set
of complex coupled ordinary differential equations. The number of ordinary
differential equations is determined by how many bound states the B-spline
method is able to find. In the cases analyzed from now on we considered up to
thirty bound states. The numerical integration of the ordinary differential
equation was performed using standard Runge-Kutta algorithms.

Before
analyzing the behavior of the exciton time-evolution it is worth to remark
that the model allows only one exciton but the electron-hole pair can occupy
many different exciton eigenstates and not only those associated to the qubit.
Also, the model does not allow the ``ionization'' of the quantum dot nor the
spontaneous recombination of the electron-hole pair in the time scale
associated to the periodic external driving.

As $|U_0|^2$ and $|U_1|^2$ are the probability that the exciton is in state
$|0\rangle$ or in state $|1\rangle$ respectively,
we use the {\em leakage}, $L$, to characterize the probability loss that
experiments the qubit when the driving  $E(t) = E_0 \sin(\omega t )$ is
applied. The leakage is defined as
\begin{equation}
 L = \lim_{n \rightarrow \infty} \frac{1}{nT} \int_{t^{\prime}}^{t^{\prime}+nT}
\;  \left( 1 - |U_0|^2 - |U_1|^2\right) \, dt \; ,
\end{equation}
where $T = 2\pi/\omega$.
\begin{figure}
\begin{center}
\includegraphics[scale=0.3]{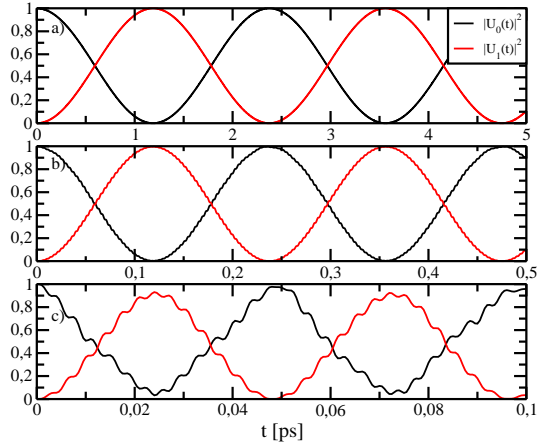}
\end{center}
\caption{\label{fig:time-evolution}(color on-line) The time evolution of the
occupation probabilities $|U_0|^2$ (black solid line) and $|U_1|^2$ (red solid
line). The exciton is initialized, in all cases, in the fundamental state, so
$|U_0(t=0)|^2 = 1$, and $|U_i(t=0)|^2=0$, $\forall i \neq 0$. From top to
bottom, panel a) shows the time evolution for $E_0 = 1\times 10^{-3}$ eV/nm,
panel b) for $E_0 = 1\times 10^{-2}$eV/nm and panel c) for $E_0 = 5\times
10^{-2}$eV/nm, respectively.  }
\end{figure}

Form now on, we consider a CdS/HgS/CdS structure with $a=b/2$. 
Figure~\ref{fig:time-evolution} shows the behavior of the occupation
probabilities $|U_0|^2$ and $|U_1|^2$ as a function of time, for different
external field strengths $E_0$. In the three cases shown, the frequency of the
external driving is set equal to the resonance frequency $\omega_{res}$ of the
exciton ground state. The resonance frequency, $\omega_{res} = (E_0 +
E_{g1})/\hbar$, where $E_0$ is the lowest eigenvalue calculated from
the exciton Hamiltonian and $E_{g1}$ is the energy gap of material one, {\em
i.e.} $E_{g1} = E^{gap}_{HgS}$.

\begin{figure}
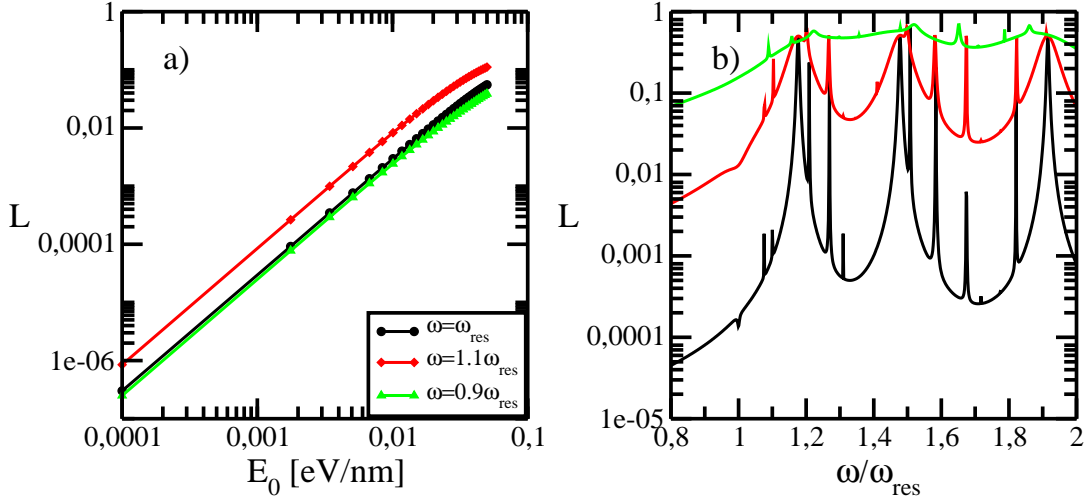

\begin{center}
\includegraphics[scale=0.5]{fig11.eps}
\includegraphics[scale=0.5]{fig12.eps}
\end{center}
\caption{\label{fig:leakage-vs-w}(color on-line) a) The {\em leakage}  of
probability {\em vs} the strength of the driving field. The linear dependency of
$L$ with $E_0$, when the axis scale is log-log, can be clearly appreciated.
From top to bottom, the Figure shows curves obtained for $\omega=0.9
\omega_{res}$ (green line and symbols), $\omega= \omega_{res}$ (black line and
symbols) and $\omega=1.1 \omega_{res}$. The leakage for $\omega=0.9
\omega_{res}$ is smaller than for $\omega = \omega_{res}$ because the qubit
does not leaves the fundamental state. b) The leakage as a function of the
external driving frequency for three different values of $E_0$. Here we use
the same external driving strengths that in Figure~\ref{fig:time-evolution}.
From bottom to top, $E_0 = 1\times 10^{-3}$ eV/nm (black line),
 $E_0 = 1\times 10^{-2}$eV/nm (red line) and  $E_0 = 5\times
10^{-2}$eV/nm (green line).}
\end{figure}

From the different panels in Figure~\ref{fig:time-evolution}, it is
possible to
appreciate that switching times on the order of picoseconds or less are
achievable for driving strengths small with no noticeable leakage. This scenario
is further supported by the data shown in Figure~\ref{fig:leakage-vs-w}.

Figure~\ref{fig:leakage-vs-w}a) shows the {\em leakage} as a function of the
strength of the external driving $E_0$ for several driving frequencies $\omega$.
The data is shown in a $\log-\log$
scale, and under this assumption the leakage shows a linear behavior for a span
of a few magnitude orders. The different curves correspond to different values
of the driving frequency, but to analyze the dependency of the leakage with the
driving frequency we choose to plot it at fixed values of the driving strength.
Figure~\ref{fig:leakage-vs-w} shoes the behavior of the leakage as a function of
the external drving frequency and for the three external driving strengths used
in Figure~\ref{fig:time-evolution}. The different curves show a rich structure,
and several spikes which are present in all the curves for the same
frequencies. These spikes are owed to the presence of many one-exciton levels
that are very close to the ground state energy. It is clear that to obtain low
levels of leakage, besides an excellent tuning of the driving frequency, it
is necessary to have well resolved exciton energy levels or, in other words,
the nano-structure should be designed in such way that the exciton is formed in
the non-perturbative regime, {\i.e} when the binding energy is as large as
possible. This seems to advice the use of large potential wells, but there is a
trade off to consider since the separation between the one-particle levels
diminishes when the characteristic sizes of the potential well are increased. A
simple way to enhance the interaction between electron-hole pair, at least in
principle, is to choose materials whose combination provides a deep potential
well and a low potential-well dielectric constant.

\section{Conclusions and Discussion}
\label{section:conclusions}

One of the advantages of using the B-spline method is its adaptability to
tackle problems with step-like parameters and matching conditions at the
interface between different spatial regions. Similarly, the method allows to
tackle complicated one or two-particle potentials, with a limited number of
adaptations. The interaction potential between the electron and the hole,
Equation~\ref{eq:electrostatic-potential} can be treated modifying the method
usually employed with the Coulomb potential \cite{deboor}, since both can be
written as expansions in spherical harmonics. 

As the analysis of the von Neumann entropy shows, for large QD (or small
cores in our case) the perturbation theory calculations give a rather poor
approximation for the binding energy. We can be fairly sure that our results,
since they are variational, that predict smaller ground state energy for the
exciton Hamiltonian, Equation~\ref{eq:hamiltoniano-exciton}, than other methods
are more accurate than previous results. This implies that our results 
predict larger values for the exciton binding energy.

To avoid an excessive leakage of probability, it is mandatory to design a
quantum dot such that the two lower states of the exciton are well
separated from the other one-exciton states. Choosing materials that allow for
a larger interaction between the electron and the hole seems to be an apparent
solution.

\acknowledgments
We would like to acknowledge  SECYT-UNC,  and CONICET 
for partial financial support of this project. We acknowledge
the fruitful discussions with Dr. C\'esar Proetto in the early stages of this
work.

\end{document}